\documentclass[twocolumn,aps,showpacs,amsmath,amssymb,floatfix,superscriptaddress]{revtex4}
\usepackage{graphicx}
\usepackage{dcolumn}
\usepackage{bm}
\usepackage{hyperref}
\usepackage{latexsym}
\usepackage{float}
\usepackage{supertabular}
\usepackage{longtable}

\begin{document}
\title{Finding Scientific Gems with Google}
\author{P.~Chen}
\email{patrick@bu.edu}
\affiliation{Center for Polymer Studies and
and Department of Physics, Boston University, Boston, MA, 02215}
\author{H.~Xie}
\email{hxie@bnl.gov}
\affiliation{New Media Lab, The Graduate Center, CUNY, New York, NY, 10016}
\affiliation{Department of Condensed Matter Physics and Materials Science,
Brookhaven National Laboratory, Upton, NY, 11973}
\author{S.~Maslov}
\email{maslov@bnl.gov}
\affiliation{Department of Condensed Matter Physics and Materials Science,
Brookhaven National Laboratory, Upton, NY, 11973}
\author{S.~Redner}
\email{redner@bu.edu}
\affiliation{Center for Polymer Studies and
and Department of Physics, Boston University, Boston, MA, 02215}

\begin{abstract}

  We apply the Google PageRank algorithm to assess the relative importance of
  all publications in the Physical Review family of journals from 1893--2003.
  While the Google number and the number of citations for each publication
  are positively correlated, outliers from this linear relation identify
  some exceptional papers or ``gems'' that are universally familiar to physicists.

\end{abstract}


\maketitle

\section{INTRODUCTION}

With the availability of electronically available citation data, it is now
possible to undertake comprehensive studies of citations that were
unimaginable just a few years ago.  In most previous studies of citation
statistics, the metric used to quantify the importance of a paper is its
number of citations.  In terms of the underlying citation network, in which
nodes represent publications and directed links represent a citation from a
{\em citing} article to a {\em cited} article, the number of citations to an
article translates to the in-degree of the corresponding node.  The
distribution of in-degree for various citation data sets has a broad tail
\cite{P65} that is reasonably approximated by a power law \cite{LS98,R98,PT}.

While the number of citations is a natural measure of the impact of a
publication, we probably all have encountered examples where citations do not
seem to provide a full picture of the influence of a publication.  We are
thus motivated to study alternative metrics that might yield a truer measure
of importance than citations alone.  Such a metric already exists in the form
of the Google PageRank algorithm \cite{Google}.  A variant of the PageRank
algorithm was recently applied to better calibrate the impact factor of
scientific journals \cite{Impact}.

In this work, we apply Google PageRank to the Physical Review citation
network with the goal of measuring the importance of individual scientific
publications published in the APS journals.  This network consists of 353,268
nodes that represent all articles published in the Physical Review family of
journals from the start of publication in 1893 until June 2003, and 3,110,839
links that represent all citations {\em to} Physical Review articles {\em
  from} other Physical Review articles.  As found previously \cite{PT}, these
internal citations represent $1/5$ to $1/3$ of all citations for highly-cited
papers.  This range provides a sense of the degree of completeness of the
Physical Review citation network.

With the Google PageRank approach, we find a number of papers with a modest
number of citations that stand out as exceptional according to the Google
\mbox{PageRank} analysis.  These exceptional publications, or gems, are
familiar to almost all physicists because of the very influential contents of
these articles.  Thus the Google PageRank algorithm seems to provide a new
and useful measure of scientific quality.

\section{THE GOOGLE PAGERANK ALGORITHM}

To set the stage for our use of Google PageRank to find scientific gems, let
us review the elements of the PageRank algorithm.  Given a network of $N$
nodes $i=1,2,\ldots,N$, with directed links that represent references from an
initial (citing) node to a target (cited) node, the Google number $G_i$ for
the $i^{\rm th}$ node is defined by the recursion formula \cite{Google}:
\begin{equation}
\label{eq-gr}
G_i = (1-d)\sum_{j\ {\rm nn\ } i}\frac{G_j}{k_j} + \frac{d}{N} \,.
\end{equation}
Here the sum is over the neighboring nodes $j$ in which a link points to node
$i$.  The first term describes propagation of the probability distribution of
a random walk in which a walk at node $j$ propagates to node $i$ with
probability $1/k_j$, where $k_j$ is the out-degree of node $j$.  The second
term describes the uniform injection of probability into the network in which
each node receives a contribution $d/N$ at each step.

Here $d$ is a free parameter that controls the performance of the Google
PageRank algorithm.  The prefactor $(1-d)$ in the first term gives the
fraction of random walks that continue to propagate along the links; a
complementary fraction $d$ is uniformly re-injected into the network, as
embodied by the second term.

We suggest that the Google number $G_i$ of paper $i$, defined by
Eq.~\eqref{eq-gr}, is a better measure of importance than the number of
citations alone in two aspects: i) being cited by influential papers
contributes more to the Google number than being cited by unimportant papers;
ii) being cited by a paper that itself has few references gives a larger
contribution to the Google number than being cited by a paper with hundreds
of references.  The Google number of a paper can be viewed as a measure of
its influence that is then equally exported to all of its references.  The
parameter $d>0$ prevents all of the influence from concentrating on the
oldest papers.

\begin{figure}[ht]
 \vspace*{0.cm}
  \includegraphics*[width=0.125\textwidth]{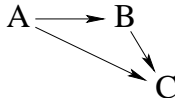}
  \caption{Feed-forward citation loop: Publication A cites both publications
    B and C.  About 50\% of entries B in the reference list of a typical
    publication A cite at least one other article C in the same reference
    list.}
\label{feed-forward}
\end{figure}

In the original Google PageRank algorithm of Brin and Page \cite{Google}, the
parameter $d$ was chosen to be 0.15.  This value was prompted by the
anecdotal observation that an individual surfing the web will typically
follow of the order of 6 hyperlinks, corresponding to a leakage probability
$d=1/6 \simeq 0.15$, before becoming either bored or frustrated with this
line of search and beginning a new search.  In the context of citations, we
conjecture that entries in the reference list of a typical paper are
collected following somewhat shorter paths of average length 2, making the
choice $d=0.5$ more appropriate for a similar algorithm applied to the
citation network.  The empirical observation justifying this choice is that
approximately 50\% of the articles \cite{caveat} in the reference list of a
given paper A have at least one citation B $\to$ C to another article C that
is also in the reference list of A (Fig.~\ref{feed-forward}).  Assuming that
such ``feed-forward'' loops result from authors of paper A following
references of paper B, we estimate the probability $1-d$ to follow this
indirect citation path to be close to 0.5.

To implement the Google PageRank algorithm for the citation network, we start
with a uniform probability density equal to $1/N$ at each node of the network
and then iterate Eq.~\eqref{eq-gr}.  Eventually a steady state set of Google
numbers for each node of the network is reached.  These represent the
occupation probabilities at each node for the random-walk-like process
defined by Eq.~\eqref{eq-gr}.  Finally, we sort the nodal Google numbers to
determine the Google rank of each node.  It is both informative and
entertaining to compare the Google rank with the citation (in-degree) rank of
typical and the most important publications in Physical Review.

\section{GOOGLE PAGERANK FOR PHYSICAL REVIEW}

\begin{figure}[ht]
 \vspace*{0.cm}
  \includegraphics*[width=0.45\textwidth]{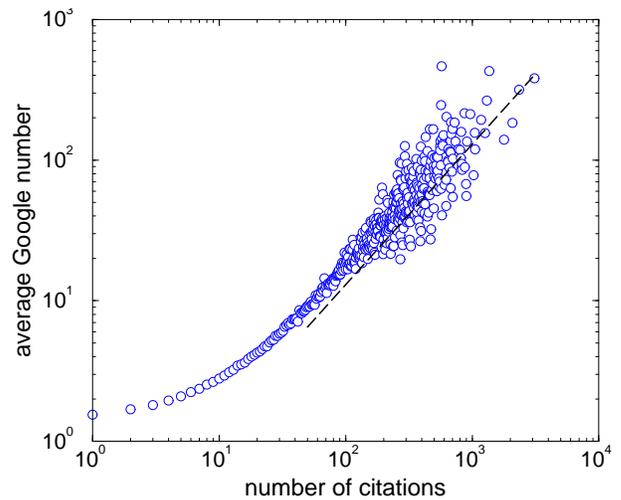}
  \caption{Average Google number $\langle G(k)\rangle$ versus number of
    citations $k$.  The dashed line of slope 1 is a guide for the eye.}
\label{g-vs-c}
\end{figure}

\begin{figure}[ht]
 \vspace*{0.cm}
  \includegraphics*[width=0.45\textwidth]{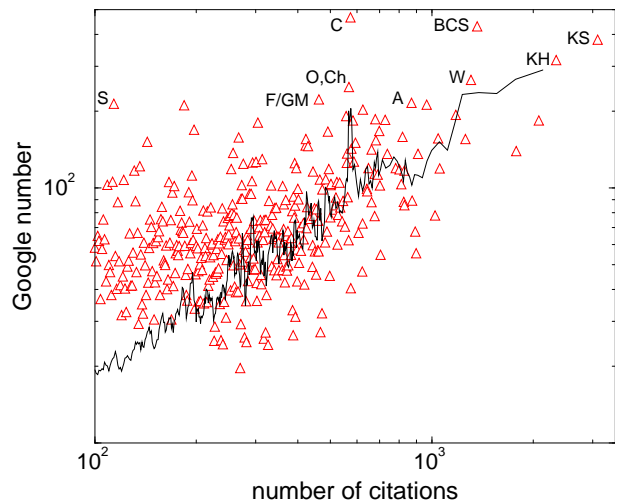}
  \caption{Individual outlier publications.  For each number of citations
    $k$, the publication with the highest Google number is plotted.  The
    top-10 Google-ranked papers are identified by author(s) initials (see
    Table~\ref{top-10}).  As a guide to the eye, the solid curve is a 5-point
    average of the data of $\langle G(k)\rangle$ versus $k$ in
    Fig.~\ref{g-vs-c}.}
\label{extreme}
\end{figure}

Fig.~\ref{g-vs-c} shows the average Google number $\langle G(k)\rangle$ for
publications with $k$ citations as a function of $k$.  For small $k$, there
are many publications with the same number of citations and the dispersion in
$G(k)$ is small.  Correspondingly, the plot of $\langle G(k)\rangle$ versus
$k$ is smooth and increases linearly with $k$ for $k\agt 50$.  Thus the
average Google number and the number of citations represent similar measures
of popularity, a result that has been observed previously \cite{FBFM,FFM}.
In fact, the citation and Google number distributions are qualitatively
similar, further indicating that citations and Google numbers are, on the
average, similar measures of importance.

However, for large $k$, much more interesting behavior occurs.  When $k$ is
sufficiently large, there is typically only one publication with $k$
citations.  Thus instead of an average value, Fig.~\ref{extreme} shows the
individual publications with the largest Google number for each value of
citation number when $k\geq 100$.  Of particular interest are the extreme
outliers with respect to the linear behavior of Fig.~\ref{g-vs-c}.  The ten
articles with the highest Google numbers are shown explicitly and are
identified by author initials (see Table~\ref{top-10}).  Also given in
Table~\ref{top-10} is the number of citations and the citation rank of these
publications.  While several of the highest-cited Physical Review papers
appear on this list, there are also several more modestly-cited papers that
are highly ranked according to the Google algorithm.

The disparity between the the Google rank and citation rank arises because,
as mentioned in the previous section, the former involves both the in-degree
as well as the Google PageRank of the neighboring nodes.  According to the
Google algorithm of Eq.~\eqref{eq-gr}, a citing publication (``child'') $j$
contributes a factor $\langle G_j/k_j\rangle$ to the Google number of its
parent paper $i$.  Thus for a paper to have a large Google number, its
children should be important (large $G_j$), and also each child should have a
small number of parents (small out-degree $k_j$).  The latter ensures that
the Google contribution of a child is not strongly diluted.

With this perspective, let us compare the statistical measures of the two
articles ``Unitary Symmetry and Leptonic Decays'', Phys.\ Rev.\ Lett.\ {\bf
  10}, 531 (1963) by N. Cabibbo (C) and ``Self-Consistent Equations Including
Exchange and Correlation Effects'', Phys.\ Rev.\ {\bf 140}, A1133 (1965) by
W. Kohn \& L. J. Sham (KS).  The former has the highest Google number of all
Physical Review publications, while the latter is the most cited.  The high
Google rank of C stems from the fact that that value of $\langle
G_j/k_j\rangle =1.52\times10^{-6}$ for the children of C is an order of
magnitude larger than the corresponding value $\langle G_j/k_j\rangle
=2.31\times10^{-7}$ for the children of KS.  This difference more than
compensates for the factor 5.6 difference in the number of citations to these
two articles (3227 for KS and 574 for C as of June 2003).  Looking a little
deeper, the difference in $\langle G_j/k_j\rangle$ for C and KS stems from
the denominator; the children of C have 15.6 citations an average, while the
children of KS are slightly ``better'' and have 18.4 citations on average.
However, the typical child of C has fewer references than a child of KS and a
correspondingly larger contribution to the Google number of C.

\begin{widetext}
{\small\begin{longtable}{|>{\hfill}p{0.39in}|c|>{\hfill}p{0.31in}|>{\hfill}p{0.27in}|p{0.28in}|>{\hfill}p{0.2in}|>{\hfill}p{0.375in}|%
>{\hfill}p{0.3in}|p{1.9in}|p{1.95in}|}
\caption{The top 10 Google-ranked publications when $d=0.5$}\label{top-10}\\
\hline
Google & Google \# & cite  &\#~~ & \multicolumn{4}{|c|}{}             &       & \\
rank& ($\times 10^{-4}$)  &rank   &  cites &  \multicolumn{4}{|c|}{Publication} &  Title & Author(s) \\ \hline
\endfirsthead
\caption{continued from previous page}\\
\hline
 Google   &Google   &cite &\#~~ & \multicolumn{4}{|c|}{}              &       & \\
 rank& number  &rank   &  cites & \multicolumn{4}{|c|}{Publication} & Title & Author(s) \\ \hline
\endhead

1&$4.65$&54&574&  PRL& 10& 531& 1963&            Unitary Symmetry and  Leptonic... & N. Cabibbo\\ \hline
2&$4.29$&5&1364&  PR& 108& 1175& 1957&       Theory of  Superconductivity&  J. Bardeen, L. Cooper, J. Schrieffer\\ \hline
3&$3.81$&1&3227&  PR& 140& A1133& 1965&       Self-Consistent Equations...  & W. Kohn \& L. J. Sham \\ \hline
4&$3.17$&2&2460&  PR& 136& B864& 1964&        Inhomogeneous Electron Gas&  P. Hohenberg \& W. Kohn \\ \hline
5&$2.65$&6&1306&  PRL& 19& 1264& 1967&         A Model of Leptons&  S. Weinberg\\ \hline
6&$2.48$&55&568&  PR& 65& 117& 1944&            Crystal Statistics&  L. Onsager\\ \hline
7&$2.43$&56&568&  RMP& 15& 1& 1943&               Stochastic Problems in...   & S. Chandrasekhar\\ \hline
8&$2.23$&95&462&     PR& 109& 193& 1958&        Theory of the Fermi     Interaction   &R. P. Feynman \& M. Gell-Mann\\ \hline
9&$2.15$&17&871&  PR& 109& 1492& 1958&           Absence of Diffusion in...  &  P. W. Anderson\\ \hline
10&$2.13$&1853&114&  PR& 34& 1293& 1929&             The Theory of Complex Spectra  &  J. C. Slater \\ \hline
\end{longtable}
}
\end{widetext}

The remaining research articles on the top-10 Google-rank list but outside
the top-10 citation list are easily recognizable as seminal publications.
For example, Onsager's 1944 paper presents the exact solution of the
two-dimensional Ising model; both a calculational {\em tour de force}, as
well as a central development in the theory of critical phenomena. The paper
by Feynman and Gell-Mann introduced the $V-A$ theory of weak interactions
that incorporated parity non-conservation and became the ``standard model''
of weak interactions.  Anderson's paper, ``Absence of Diffusion in Certain
Random Lattices'' gave birth to the field of localization and is cited by the
Nobel prize committee for the 1977 Nobel prize in physics.

The last entry in the top-10 Google-rank list, ``The Theory of Complex
Spectra'', by J. C.  Slater (S) is particularly striking.  This article has
relatively few citations (114 as of June 2003) and a relatively low citation
rank ($1853^{\rm th}$), but its Google number $2.13\times 10^{-4}$ is only a
factor 2.2 smaller than that of Cabibbo's paper!  What accounts for this high
Google rank?  From the scientific standpoint, Slater's paper introduced the
determinant form for the many-body wavefunction.  This form is so ubiquitous
in current literature that very few articles actually cite the original work
when the Slater determinant is used.  The Google PageRank algorithm
identifies this hidden gem primarily because the average Google contribution
of the children of S is $\langle G_j/k_j\rangle =3.51\times10^{-6}$, which is
a factor 2.3 larger than the contribution of the children of C.  That is, the
children of Slater's paper were both influential and Slater loomed as a very
important father figure to his children.

\begin{widetext}
{\small\begin{longtable}{|>{\hfill}p{0.39in}|c|>{\hfill}p{0.39in}|>{\hfill}p{0.27in}|p{0.28in}|>{\hfill}p{0.2in}|>{\hfill}p{0.32in}|%
>{\hfill}p{0.3in}|p{1.9in}|p{1.9in}|}
\caption{The remaining top-100 Google-ranked papers when $d=0.5$ in which the
  ratio of Google rank to citation rank is greater than 10.}\label{tab-hprlcr}\\
\hline
Google & Google \# & cite  &\#~~ & \multicolumn{4}{|c|}{}             &       & \\
rank& ($\times 10^{-4}$) &rank   &  cites &  \multicolumn{4}{|c|}{Publication} &  Title & Author(s) \\ \hline
\endfirsthead
\caption{continued from previous page}\\
\hline
 Google   &Google   &cite &\#~~ & \multicolumn{4}{|c|}{}              &       & \\
 rank& \# ($\times 10^{-4}$) &rank   &  cites & \multicolumn{4}{|c|}{Publication} & Title & Author(s) \\ \hline
\endhead
1&$4.65$&54&574&  PRL& 10& 531& 1963&            Unitary Symmetry and
  Leptonic... & N. Cabibbo\\ \hline
8&$2.23$&95&462&     PR& 109& 193& 1958&        Theory of the Fermi
     Interaction   &R. P. Feynman \& M. Gell-Mann\\ \hline
10&$2.13$&1853&114&  PR& 34& 1293& 1929&             The Theory of Complex Spectra
  &  J. C. Slater \\ \hline
12&$2.11$&712&186&  PR0& 43& 804& 1933& On the Constitution of$\ldots$ & E. Wigner \& F. Seitz\\ \hline
20&$1.80$&228&308&  PR0& 106& 364& 1957& Correlation Energy of an $\ldots$ &
M. Gell-Mann \& K. Brueckner\\ \hline
21&$1.69$&616&198&  PRL& 58& 408& 1987& Bulk superconductivity at $\ldots$ &
R. J. Cava et al. \\ \hline
25&$1.58$&311&271&  PRL& 58& 405& 1987& Evidence for superconductivity $\ldots$ & C. W. Chu et al.\\ \hline
30&$1.51$&1193&144&  PRL& 10& 84& 1963& Photon Correlations & R. J. Glauber\\ \hline
35&$1.42$&12897&39&  PR0& 35& 509& 1930& Cohesion in Monovalent Metals & J. C. Slater\\ \hline
49&$1.21$&1342&136&  PR0& 60& 252& 1941& Statistics of the Two- $\ldots$ & H. A. Kramers \& G. H. Wannier\\ \hline
58&$1.17$&1433&135&  PR0& 81& 440& 1951& Interaction Between the $\ldots$ & C. Zener\\ \hline
59&$1.17$&5196&66&  PR0& 45& 794& 1934& Electronic Energy Bands in $\ldots$ & J. C. Slater\\ \hline
60&$1.16$&2927&108&  PRB& 28& 4227& 1983& Electronic structure of $\ldots$ & L. F. Mattheiss \& D. R. Hamann\\ \hline
64&$1.12$&642&199&  PR0& 52& 191& 1937& The Structure of Electronic $\ldots$ & G. H. Wannier\\ \hline
70&$1.08$&1653&130&  PRL& 10& 518& 1963& Classification of Two-Electron $\ldots$ & J. Cooper, U. Fano \& F. Prats\\ \hline
72&$1.06$&1901&118&  PR0& 46& 509& 1934& On the Constitution of $\ldots$ & E. Wigner \& F. Seitz\\ \hline
73&$1.05$&876&180&  PR0& 75& 486& 1949& The Radiation Theories of $\ldots$ & F. J. Dyson\\ \hline
78&$1.03$&1995&119&  PR0& 109& 1860& 1958& Chirality Invariance and $\ldots$ & E. Sudarshan \& R. Marshak\\ \hline
85&$1.00$&201853&3&  PRB& 22& 5797& 1980& Cluster formation in $\ldots$ & H. Rosenstock \& C. Marquardt\\ \hline
87&$0.99$&10168&48&  PRL& 6& 106& 1961& Population Inversion and $\ldots$ &
A. Javan, W. Bennett, D. Herriott\\ \hline
90&$0.98$&3231&86&  PR0& 79& 350& 1950& Antiferromagnetism. $\ldots$ & P. W. Anderson\\ \hline
92&$0.97$&1199&149&  PR0& 76& 749& 1949& The Theory of Positrons & R. P. Feynman\\ \hline
\end{longtable}
}
\end{widetext}

The striking ability of the Google PageRank algorithm to identify influential
papers can be seen when we consider the top-100 Google-ranked papers.
Table~\ref{tab-hprlcr} shows the subset of publications on the top-100 Google
rank in which the ratio of Google rank to citation rank is greater than 10;
that is, publications with anomalously high Google rank compared to their
citation rank.  This list contains many easily-recognizable papers for the
average physicist.  For example, the publication by Wigner and Seitz, ``On
the Constitution of Metallic Sodium'' introduced Wigner-Seitz cells, a
construction that appears in any solid-state physics text.  The paper by
Gell-Mann and Brueckner, ``Correlation Energy of an Electron Gas at High
Density'' is a seminal publication in many-body theory.  The publication by
Glauber, ``Photon Correlations'', was recognized for the 2005 Nobel prize in
physics.  The Kramers-Wannier article, ``Statistics of the Two-Dimensional
Ferromagnet. Part I'', showed that a phase transition occurs in two
dimensions, contradicting the common belief at the time.  The article by
Dyson, ``The Radiation Theories of Tomonaga, Schwinger, and Feynman'',
unified the leading formalisms for quantum electrodynamics and it is
plausible that this publication would have earned Dyson the Nobel prize if it
could have been shared among four individuals.  One can offer similar
rationalizations for the remaining articles in this table.

On the other hand, an apparent mistake is the paper by Rosenstock and
Marquardt, ``Cluster formation in two-dimensional random walks: Application
to photolysis of silver halides'' (RM).  Notice that this article has only 3
citations!  Why does RM appear among the top-100 Google-ranked publications?
In RM, a model that is essentially diffusion-limited aggregation is
introduced.  Although these authors had stumbled upon a now-famous model,
they focused on the kinetics of the system and apparently did not appreciate
its wonderful geometrical features.  This discovery was left to one of the
children of RM---the famous paper by T.  Witten and L.  Sander,
``Diffusion-Limited Aggregation, a Kinetic Critical Phenomenon'' Phys.\ Rev.\
Lett.\ {\bf 47}, 1400 (1981), with 680 citations as of June 2003.
Furthermore, the Witten and Sander article has only 10 references; thus a
substantial fraction of its fame is exported to RM by the Google PageRank
algorithm.  The appearance of RM on the list of top-100 Google-ranked papers
occurs precisely because of the mechanics of the Google PageRank algorithm in
which being one of the few references of a famous paper makes a huge
contribution to the Google number.

\begin{figure}[ht]
 \vspace*{0.cm}
  \includegraphics*[width=0.5\textwidth]{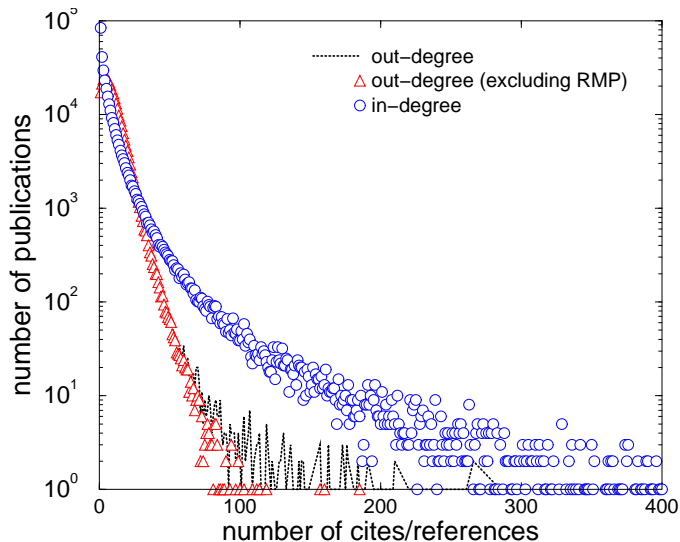}
  \caption{The in-degree distribution (citations to) and out-degree
    distribution (references from) for all Physical Review publications.  The
    out-degree distribution is shown with and without the contribution of
    Reviews of Modern Physics.}
\label{in-out}
\end{figure}

A natural question to ask is whether the Google rankings are robust with
respect to the value of the free parameter $d$ in the Google algorithm.  As
mentioned above, we believe that our {\it ad hoc} choice of $d=0.5$ accounts
in a reasonable way for the manner in which citations are actually made.  For
$d=0.15$, as in the original Google algorithm, the Google rankings of
highly-cited papers locally reorder to a considerable extent compared to the
rankings for the case $d=0.5$, but there is little global reordering.  For
example, all of the top 10 Google-ranked papers calculated with $d=0.5$
remained among the top-50 Google-ranked papers for $d=0.15$.  Thus up to this
degree of variation, Google rankings are a robust measure.

On the other hand, in the limit $d\to 1$ Google rank approaches
citation rank.  For example, for $d=0.9$, 7 of the top-10 Google-ranked
papers with $d=0.9$ are also among the 10 most-cited particles, while the
citation ranks of the remaining 3 of the top-10 Google-ranked articles are
19, 54, and 56.  In fact, we argue that the Google rank reduces to citation
rank as $d\to 1$.  To show this, we first notice that in the extreme case of
$d=1$, the Google number of each node equals $1/N$.  For $d\to 1$, we
therefore write $d=1-\epsilon$, with $\epsilon \ll 1$, and also assume that
there is a correspondingly small deviation of the Google numbers from $1/N$.
Thus we write $G_i=\frac{1}{N}+\mathcal{O}(\epsilon)$.  Substituting these
into Eq.~\eqref{eq-gr}, we obtain
\begin{eqnarray}
\label{Gi}
G_i &=& \epsilon\sum_{j} \frac{G_j}{k_j} + \frac{1-\epsilon}{N}  \nonumber\\
&\approx& \frac{1}{N}\left[1+\epsilon\left(\sum_{j} \frac{1}{k_j} -1\right)\right]
\end{eqnarray}
To estimate the sum in Eq.~\eqref{Gi}, we use the fact that the out-degree
distribution is relatively narrow (Fig.~\ref{in-out}), especially if we
exclude the broad tail that is caused by the contributions of review articles
that appear in the Reviews of Modern Physics.  While the mean in-degree and
out-degrees are both close to 9 (and should be exactly equal for the complete
citation network), the dispersion for the in degree is 23.15, while the
dispersion for the out degree (excluding Reviews of Modern Physics) is 8.64.

As a result of the sharpness of the out-degree distribution, the sum
$\sum_{j\ {\rm nn\ } i} \frac{1}{k_j}$ for nodes with high in-degree is
approximately equal to the in-degree $d_i$ of node $i$ times $\langle
\frac{1}{k}\rangle$.  With this assumption, Eq.~\eqref{Gi} becomes
\begin{equation}
\label{eqn_d1}
G_i = \frac{1}{N}\left[1+\epsilon\left(d_i\left\langle
\frac{1}{k}\right\rangle -1\right)\right]\,.
\end{equation}
That is, the leading correction to the limiting $d=1$ result that
$G_i=\frac{1}{N}$ is proportional to the in-degree of each node.  Thus as
$d\to 1$, the Google rank of each node is identical to its citation rank
under the approximation that we neglect the effect of the dispersion of the
out-degree in the citation network.

\section{CONCLUSIONS}

We believe that protocols based on the Google \mbox{PageRank} algorithm hold
a great promise for quantifying the impact of scientific publications.  They
provide a meaningful extension to traditionally-used importance measures,
such as the number of citation of individual articles and the impact factor
for journals as a whole.  The PageRank algorithm implements, in an extremely
simple way, the reasonable notion that citations from more important
publications should contribute more to the rank of the cited paper than those
from less important ones.  Other ways of attributing a quality for a citation
would require much more detailed contextual information about the citation
itself.

The situation in citation networks is not that dissimilar from that in the
World Wide Web, where hyperlinks contained in popular websites and pointing
to your webpage would bring more Internet traffic to you and thus would
contribute substantially to the popularity of your own webpage.  Scientists
commonly discover relevant publications by simply following chains of
citation links from other papers. Thus it is reasonable to assume that the
popularity or ``citability'' of papers may be well approximated by the random
surfer model that underlies the PageRank algorithm.  One meaningful
difference between the WWW and citation networks is that citation links
cannot be updated after publication, while WWW hyperlinks keep evolving
together with the webpage containing them.  Thus scientific papers and their
citations tend to age much more rapidly than active webpages.  These
differences could be taking into account by explicitly incorporating the
effects of aging into the Page Rank algorithm \cite{Aging_google}.

\acknowledgments{Two of us, (PC and SR) gratefully acknowledge financial
  support from the US National Science Foundation grant DMR0535503.  SR also
  thanks A. Castro Neto, A. Cohen, and K. Lane for literature advice.  Work
  at Brookhaven National Laboratory was carried out under Contract No.\
  DE-AC02-98CH10886, Division of Material Science, U.S. Department of Energy.
}


\begin{thebibliography}{99}
\bibitem{caveat} The actual fraction of "followed citations" (such as B in
  Fig.~\ref{feed-forward}) is 42\% for the entire dataset and 51\% for papers
  published during the last 4 years.
\end{thebibliography}
\end{document}